\documentclass[aps,prd,preprintnumbers,nofootinbib,floatfix,11pt]{revtex4}

\usepackage{amsfonts,amsmath,amssymb,mathrsfs,graphicx,epsfig,color,amsthm} 

\allowdisplaybreaks[1]
\newcommand{\be}{\begin{equation}}
\newcommand{\ee}{\end{equation}}
\newcommand{\ba}{\begin{eqnarray}}
\newcommand{\ea}{\end{eqnarray}}

\theoremstyle{definition}
\newtheorem{theorem}{Theorem}

\newtheorem{definition}[theorem]{Definition}

\usepackage{}

\usepackage[normalem]{ulem}

\begin{document}

\title{Attractive gravity probe surface in Einstein-Maxwell system}
\author{Kangjae Lee${}^1$,  Keisuke Izumi$^{1,2}$, Tetsuya Shiromizu$^{1,2}$, Hirotaka Yoshino$^3$ and Yoshimune Tomikawa$^4$}

\affiliation{$^{1}$Department of Mathematics, Nagoya University, Nagoya 464-8602, Japan}
\affiliation{$^{2}$Kobayashi-Maskawa Institute, Nagoya University, Nagoya 464-8602, Japan} 
\affiliation{$^{3}$Department of Physics, Osaka Metropolitan University, Osaka 558-8585, Japan}
\affiliation{$^{4}$Division of Science, School of Science and Engineering, Tokyo Denki University, Saitama 350-0394, Japan}

\begin{abstract}
\begin{center}
{\bf Abstract}
\end{center}
\noindent
We derive areal inequalities
for five types of attractive gravity probe surfaces, which
were proposed by us in order to 
characterize the strength of gravity in different ways including weak gravity
region, taking into account of contributions of electric and magnetic charges, angular momentum, gravitational waves, and matters.
These inequalities are generalizations of the Riemannian Penrose inequality
for minimal surfaces,
and lead to the concept of  
extremality for a given surface whose condition
is given in terms of the gravitational
mass and the electromagnetic charges.
This means that the extremality 
is a characteristic property not only of black hole horizons or minimal surfaces
but also of surfaces in weak gravity 
region. We also derive areal inequalities and extremality conditions for surfaces in asymptotically locally anti-de Sitter spacetimes.
\end{abstract}

\maketitle


\section{Introduction}

It is expected that the size of black holes has an upper bound which coincides with the Schwarzschild radius because of strong gravity, 
and the corresponding areal inequality, called the Penrose inequality~\cite{Penrose1973}, has been proposed. 
For time-symmetric spacelike hypersurfaces, the outermost trapped surface coincides with the outermost minimal surface, and then
the Penrose inequality is reduced to an inequality for the outermost minimal surface on a spacelike hypersurface with non-negative scalar curvature. 
This inequality is called the Riemannian Penrose inequality, which is given by $A\le4\pi(2m)^2$ where $A$ is the area of the minimal surface and $m$ is the Arnowitt-Deser-Misner (ADM) mass. 
This problem has been positively solved in Refs.~\cite{wald, imcf, bray}. 
The inequalities including the contributions of the electric and magnetic charges~\cite{Jang:1979zz,Weinstein2004,Disconzi:2012es,Khuri:2013dfa,Khuri:2014wqa,Khuri:2013ana} 
and the angular momentum~\cite{Cha:2014yja,Cha:2014oja,Anglada2017, Anglada2020, Khuri2018, Dain2018} have also been given under some assumptions.

Since the trapped surfaces are hidden by the event horizons, the Penrose inequality concerns the unobservable surfaces.
This motivates us to introduce
new concepts, called loosely trapped surface (LTS)~\cite{Shiromizu2017} and dynamically transversely trapping surface (DTTS)~\cite{Yoshino:2017gqv,Yoshino2019,Yoshino:2019mqw}
as generalizations of the photon sphere.
The photon sphere characterizes a strong gravity region
that exists outside the event horizon in a different way from the trapped surfaces, 
and is also an important object in the observation of black holes~\cite{EventHorizonTelescope:2019dse,EventHorizonTelescope:2022wkp,EventHorizonTelescope:2019pgp}.
The areal inequalities for the LTS and DTTS, $A_{\rm LTS/DTTS}\le4\pi(3m)^2$, have been proved~\cite{Shiromizu2017,Yoshino2019} as well.
In Ref.~\cite{Lee2020}, refined inequalities are shown
for the LTS and the DTTS taking into account of the contribution of electromagnetic fields.

Further generalization of the LTS covering weak gravity region, called an attractive gravity probe surface (AGPS) \cite{Izumi2021}, 
has been proposed, and an inequality for the area of the AGPS,
\begin{equation}
  A_{\rm AGPS}\le 4\pi m^2\left(\frac{3+4\alpha}{1+2\alpha}\right)^2,
\end{equation}
has been shown, 
where $\alpha(>-1/2)$ is a constant restricting the strength of the gravity on the surface. The existence of the AGPS implies the 
existence of positive energy~\cite{Shiromizu:2023hfs}, and an application to the cosmological horizon has also been 
discussed~\cite{Shiromizu:2023qkn}.
In Ref.~\cite{Izumi2022}, the definition of the AGPS has been refined, and the areal inequality for the refined AGPS 
has been shown in higher dimensions.
In Ref.~\cite{Lee2022b}, two kinds of generalizations
of the DTTS and one generalization of the LTS
to weak gravity region have been  
proposed in a similar manner (i.e., by introducing a parameter
to specify the strength of gravity).
These three generalizations plus the original AGPS
are classified as four types of AGPSs,
which characterize the strength of gravity on a given surface
in different manners. 
For these four types of AGPSs,
the refined areal inequalities including contributions 
of angular momentum, gravitational waves, and matters have been derived
as well~\cite{Lee2022a, Lee2022b}.

In this paper, we derive the refined areal inequalities for the four types
of AGPSs and the refined AGPS (in total, five types) 
in the Einstein-Maxwell system, 
keeping the contributions of the electromagnetic fields, angular momentum, gravitational waves and matters, 
which are usually removed as non-negative contributions in the inequalities.
In particular, the inequality including
electromagnetic contributions leads to the concept of extremality
for surfaces, 
which are analogues to that for charged black holes.
Since the inequalities and the concept of extremality would be applicable to the anti-de Sitter/conformal field theory (AdS/CFT) correspondence, 
we also analyze the cases of asymptotically locally AdS spacetimes.

This paper consists of the following sections. 
In Sec.~\ref{basics}, we give the setup and the
basic formulas used in the proofs. 
Then, we provide the definitions of the AGPSs in Sec.~\ref{Secdef}.
The inequalities for surfaces are derived in Sec.~\ref{lagps}. 
These inequalities lead to the extremality condition for AGPSs,
which is discussed in Sec.~\ref{cor}.
Asymptotically locally AdS cases are analyzed in Sec.~\ref{AdS}.
The final section, Sec.~\ref{summary}, is devoted to a summary and discussion.


\section{Setup and basic formulas}
\label{basics}

In this section, we give the setup of this paper and show the basic formulas that will be  used in the subsequent sections.

\subsection{Decomposition of spacetime}
\label{Decomp}

In a 4-dimensional spacetime ($M$, $g_{ab}$), we consider two 3-dimensional hypersurfaces; one is a spacelike hypersurface ($\Sigma$, $q_{ab}$, $K_{ab}$), and the other is a timelike hypersurface ($S$, $p_{ab}$, $\bar{K}_{ab}$).
Suppose that $\Sigma$ and $S$ are orthogonal and that their intersection is a compact spacelike 2-surface ($\sigma_0$, $h_{ab}$, $k_{ab}$, $\kappa_{ab}$).
The metric $g_{ab}$ is written by
\begin{equation}
	g_{ab}=q_{ab}-n_an_b=p_{ab}+r_ar_b=h_{ab}+r_ar_b-n_an_b,
\end{equation}
where $n^a$ is the timelike unit normal vector to $\Sigma$ and $r^a$ is the spacelike unit normal vector to $S$.
The extrinsic curvatures $K_{ab}$, $\bar{K}_{ab}$, $k_{ab}$ and $\kappa_{ab}$ are defined by 
\begin{align}
	K_{ab}&=q_a{}^cq_b{}^d \nabla_c n_d,\\
	\bar{K}_{ab}&=p_a{}^cp_b{}^d \nabla_c r_d,\\
	k_{ab}&=h_a{}^ch_b{}^d \nabla_c r_d,\\
	\kappa_{ab}&=h_a{}^ch_b{}^d \nabla_c n_d,
\end{align}
respectively, where $\nabla_a$ is the covariant derivative with respect to $g_{ab}$.
The traces of the extrinsic curvatures are denoted as $K:=K_a{}^a$, $\bar K:=\bar K_a{}^a$, $k:=k_a{}^a$ and $\kappa:=\kappa_a{}^a$.

From the geometrical identity obtained by the decomposition of spacetime dimensions, the derivatives of $k$ with respect to $r^a$ on $\Sigma$ and 
of $\kappa$ with respect to $n^a$ on $S$ can be expressed as
\begin{eqnarray}
	\label{pre.liederiv.k}
	&&r^aD_ak=-{}^{(3)}R_{ab}r^ar^b-k_{ab}k^{ab}-\varphi^{-1}{\cal D}^2\varphi, \\
	\label{pre.liederiv.kappa}
	&&n^a\bar{D}_a\kappa=-{}^{(3)}\bar{R}_{ab}n^an^b-\kappa_{ab}\kappa^{ab}+N^{-1}{\cal D}^2N,
\end{eqnarray}
where $D_a$, $\bar{D}_a$ and   ${\cal D}_a$ are the covariant derivatives on $\Sigma$, on $S$ and on $\sigma_0$, ${}^{(3)}R_{ab}$ and ${}^{(3)}\bar{R}_{ab}$ are the Ricci  tensors of  $\Sigma$ and $S$, and $\varphi$ and $N$ are the lapse functions with respect to  $r^a$ and $n^a$, respectively.
Let us decompose the extrinsic curvatures $K_{ab}$ and $\bar K_{ab}$ as 
\begin{eqnarray}
&&K_{(r)}:=K_{ab}r^ar^b , \hspace{10mm} v_a:=h_a{}^bK_{bc}r^c, \hspace{10mm} \kappa_{ab}=h_a{}^ch_b{}^dK_{cd}, \\
&&\bar K_{(n)}:= \bar K_{ab}n^a n^b , \hspace{10mm} v_a=-h_a{}^b \bar K_{bc}n^c, \hspace{10mm} k_{ab}=h_a{}^ch_b{}^d \bar K_{cd}.
\end{eqnarray}
The double traces for the Gauss equations on $\Sigma$ and on $S$ in $M$ give the following equations for
 3-dimensional scalar curvature ${}^{(3)}R$ and ${}^{(3)} \bar R$,
\begin{eqnarray}
&&	{}^{(3)}R=2G_{ab}n^an^b+2v_av^a+\tilde{\kappa}_{ab}\tilde{\kappa}^{ab}-2\kappa K_{(r)}-\frac{1}{2}\kappa^2, \\
&&	{}^{(3)} \bar R= - 2G_{ab}r^ar^b+2v_av^a-\tilde{k}_{ab}\tilde{k}^{ab}+2k \bar K_{(n)} + \frac{1}{2}k^2, 
\end{eqnarray}
where $G_{ab}$ is the 4-dimensional Einstein tensor, $\tilde \kappa_{ab}$ and $\tilde k_{ab}$ are the traceless components 
of $\kappa_{ab}$ and $k_{ab}$ respectively, that is, $\tilde \kappa_{ab}=\kappa_{ab}-(1/2)\kappa h_{ab}$ and 
$\tilde k_{ab}=k_{ab}-(1/2)k h_{ab}$. 
Using these equations, the double trace for the Gauss equation on $\sigma_0$ in $\Sigma$ and $S$, 
Eqs.~\eqref{pre.liederiv.k} and \eqref{pre.liederiv.kappa} become (see more details in Refs.~\cite{Lee2022a, Lee2022b})
\begin{equation}
	\label{liederiv.k}
	r^aD_ak=\frac{1}{2}{}^{(2)}R-G_{ab}n^an^b+\kappa K_{(r)}+\frac{1}{4}\left(\kappa^2-2 \tilde \kappa_{ab}\tilde \kappa^{ab}-3 k^2- 2\tilde k_{ab}\tilde k^{ab}\right)-v_av^a-\varphi^{-1}{\cal D}^2\varphi
\end{equation}
and
\begin{equation}
	\label{liederiv.kappa}
	n^a\bar{D}_a\kappa=-\frac{1}{2}{}^{(2)}R-G_{ab}r^ar^b-k\bar{K}_{(n)}+\frac{1}{4}\left(k^2-2\tilde k_{ab}\tilde k^{ab}-3\kappa^2-2\tilde \kappa_{ab}\tilde \kappa^{ab}\right)+v_av^a+N^{-1}{\cal D}^2N,
\end{equation}
respectively, where ${}^{(2)}R$ is the 2-dimensional scalar curvature of $\sigma_0$.

\subsection{Maxwell field}

We consider the Einstein-Maxwell system with matters. 
The total energy-momentum tensor is the sum of those of the Maxwell field and (other) matters, $T_{ab}=T^{\rm (em)}_{ab}+T^{(m)}_{ab}$. 
The energy-momentum tensor $T^{\rm (em)}_{ab}$ of the Maxwell field is given by 
\begin{equation}
	\label{EM}
	T^{\rm (em)}_{ab}=\frac{1}{4\pi}\left(F_{ac}F_b{}^c-\frac{1}{4}g_{ab}F_{cd}F^{cd}\right),
\end{equation}
where $F_{ab}$ is the field strength of the Maxwell field.
From Eq.~\eqref{EM}, we obtain
\begin{align}
	\label{em.energy}
	&8\pi T^{\rm (em)}_{ab}n^an^b=E_aE^a+B_aB^a
	=(E_ar^a)^2+(B_ar^a)^2 + 8\pi\rho^{\rm (em)}_\parallel ,\\
	\label{em.pressure}
	&8\pi T^{\rm (em)}_{ab}r^ar^b
	=-(E_ar^a)^2-(B_ar^a)^2 + 8\pi\rho^{\rm (em)}_\parallel,
\end{align}
where $E_a:=F_{ab}n^b$ and $B_a:=-{}^*F_{ab}n^b$ are the electric and magnetic fields, respectively, ${}^*F_{ab}$ is 
defined as ${}^*F_{ab}:=(1/2)\epsilon_{abcd}F^{cd}$, where $\epsilon_{abcd}$ is the Levi-Civita tensor, and $\rho^{\rm (em)}_\parallel$ is defined as 
\begin{equation}
  8\pi\rho^{\rm (em)}_\parallel:=h^{ab}(E_aE_b+B_aB_b).
  \label{Def:rho-em-parallel}
\end{equation}
For later discussion, we have decomposed the energy density into the contribution of the radial components of the electromagnetic 
fields and the other contribution of the tangential components.

For simplicity, we assume that other matter fields have no electromagnetic charges outside $\sigma_0$,
while, inside $\sigma_0$, matters can have electromagnetic charges. 
Then, we can define the total electromagnetic charges inside $\sigma_0$ as 
\begin{align}
\label{defCharge}
	4\pi q_e := \int_{\sigma_0}E_ar^adA,\hspace{10mm}
	4\pi q_m := \int_{\sigma_0}B_ar^adA.
\end{align}
Since no electromagnetic charges exist outside $\sigma_0$, these charges are conserved, meaning that the integration over a surface enclosing $\sigma_0$ 
gives the same value with $q_e$ and $q_m$. 


\subsection{Geroch monotonicity}
\label{gerochmon}

The monotonicity of the Geroch energy plays an important role in deriving our inequalities. 
Here, we briefly review it~\cite{wald,imcf,geroch}.

Suppose that the 4-dimensional spacetime ($M$, $g_{ab}$) is asymptotically flat, and $\Sigma$ is asymptotically flat spacelike hypersurface foliated by the inverse mean curvature flow (IMCF) $\{\sigma_y\}_{y\in\mathbb{R}}$ with $\sigma_y \approx S^2$. 
Geroch has shown the monotonicity of the Geroch energy%
\footnote{
Geroch monotonicity under the weak flow has been proved in Ref.~\cite{imcf}.
} defined by~\cite{geroch} 
\begin{equation}
	E(y):=\frac{A^{1/2}(\sigma_y)}{64\pi^{3/2}}\int_{\sigma_y}\left(2{}^{(2)}R-k^2\right)dA, 
\end{equation}
where $A(\sigma_y)$ is the area of $\sigma_y$. To refine the inequalities, 
we keep some non-negative contributions which are eliminated in the inequality originally derived by 
Geroch.

The IMCF is parameterized by $y$, which is defined as $r_a = \varphi D_a y$ with $\varphi k=1$. 
The derivative of Geroch energy with respect to $y$ is calculated as
\begin{equation}
	\label{pre.mon.geroch}
	\frac{dE}{dy}=\frac{A^{1/2}}{64\pi^{3/2}}\int_{\sigma_y}dA\left[{}^{(2)}R-\frac{3}{2}k^2 -2 r^aD_a k\right].
\end{equation}
Substituting Eq.~\eqref{liederiv.k} into this equation, we have
\begin{equation}
	\label{mon.geroch}
	\frac{dE}{dy}=\frac{A^{1/2}}{64\pi^{3/2}}\int_{\sigma_y}dA\left[2\varphi^{-2}({\cal D}\varphi)^2+2G_{ab}n^an^b+2v_av^a+\tilde{\kappa}_{ab}\tilde{\kappa}^{ab}+\tilde{k}_{ab}\tilde{k}^{ab}-2\kappa K_{(r)}-\frac{1}{2}\kappa^2\right],
\end{equation}
where the integration by parts was done in the first term.

Suppose that $\Sigma$ is a maximal slice
\footnote{
Even if we keep the last two terms of Eq.\eqref{mon.geroch} without assuming a maximal slice, 
it is possible to derive an inequality with additional terms to that of Eq.~\eqref{former.areaineq}. 
However, these additional terms are difficult to give physical meanings to,
and can be both positive and negative in general. 
In the discussion of the Penrose inequality, the assumption of a maximal slice
is widely used to omit these terms from the inequality. }, and then 
\begin{equation}
K=K_{(r)}+\kappa=0
\end{equation}
is satisfied. 
Through the Einstein equation $G_{ab}=8\pi T_{ab}$, $G_{ab}n^an^b$ in Eq.~\eqref{mon.geroch} can be replaced by the energy density, and we have 
\begin{eqnarray}
	\label{mon.geroch2}
	\frac{dE}{dy} &\ge & \frac{A^{1/2}}{64\pi^{3/2}}\int_{\sigma_y}dA\left[2\varphi^{-2}({\cal D}\varphi)^2+2(E_ar^a)^2+2(B_ar^a)^2+2v_av^a+16\pi (\rho^{\rm(em)}_\parallel+\rho^{(m)}+\rho_{{\rm gw}})\right] \nonumber \\
	&\ge & \frac{A^{1/2}}{32\pi^{3/2}}\int_{\sigma_y}dA\left[ (E_ar^a)^2+(B_ar^a)^2+v_av^a+8\pi (\rho^{\rm(em)}_\parallel+\rho^{(m)}+\rho_{{\rm gw}})\right],
\end{eqnarray}
where Eq.~\eqref{em.energy} was used at the first inequality. Here, $\rho^{(m)}$ is the energy density of matter fields defined 
by $\rho^{(m)}:=T_{ab}^{(m)}n^an^b$, and 
\begin{equation}
	8\pi \rho_{{\rm gw}}:=\frac{1}{2}(\tilde \kappa_{ab} \tilde \kappa^{ab}+\tilde k_{ab}\tilde k^{ab})
\end{equation} 
can be interpreted as the energy density of gravitational field.

Application of the Cauchy-Schwarz inequality to Eq.~\eqref{mon.geroch2} gives 
\begin{align}
	\label{mon.geroch3}
	\frac{dE}{dy}\ge\frac{A^{1/2}}{32\pi^{3/2}}\left[\frac{1}{A}\left(\int_{\sigma_y}E_ar^adA\right)^2+\frac{1}{A}\left(\int_{\sigma_y}B_ar^adA\right)^2+8\pi A\bar\rho+\int_{\sigma_y}dAv_av^a\right],
\end{align}
where we have introduced the area-averaged energy density $\bar \rho$ as
\begin{equation}
	\bar\rho:=\frac{1}{A(\sigma_y)}\int_{\sigma_y}dA(\rho^{\rm(em)}_\parallel+\rho^{(m)}+\rho_{{\rm gw}}).
\end{equation}

In our assumption, no source for the electric and magnetic fields exists outside $\sigma_0$ on $\Sigma$. Then, the electric and magnetic charges $q_e$ and $q_m$ defined in Eq.~\eqref{defCharge} are conserved, which implies
\begin{align}
	4\pi q_e=\int_{\sigma_y}E_ar^adA,\hspace{10mm}
	4\pi q_m=\int_{\sigma_y}B_ar^adA, 
\end{align}
for any $y$.
Equation~\eqref{mon.geroch3} is written with these charges as
\begin{align}
	\frac{dE}{dy}\ge\frac{A^{1/2}}{32\pi^{3/2}}\left[\frac{(4\pi q)^2}{A}+8\pi A\bar\rho+\int_{\sigma_y}dAv_av^a\right],
\end{align}
where $q$ is defined by $q^2:=q_e^2+q_m^2$ and is called the electromagnetic charge hereafter. 
Integrating the above inequality over $y$, we have an inequality,
\begin{align}
	\label{former.areaineq}
	m_{\rm ADM}-m_{\rm ext}-\frac{{\cal R}_{A0}}{2}+\frac{{\cal R}_{A0}}{32\pi}\int_{\sigma_0}dAk^2&\ge\frac{1}{4}\int^\infty_0dy\left(\frac{q^2}{{\cal R}_A}+6\frac{\bar J_y^2}{{\cal R}_A^3}\right)\nonumber\\
	&\ge\frac{1}{2}\frac{q^2}{{\cal R}_{A0}}+\frac{\bar J^2_{\rm min}}{{\cal R}_{A0}^3},
\end{align} 
where ${\cal R}_{A}$ and ${\cal R}_{A0}$ are the areal radius of $\sigma_y$ and $\sigma_0$, that is,
\begin{eqnarray}
	&&{\cal R}_{A}:= \sqrt{\frac{A}{4\pi}} \hspace{10mm}\mbox{with} \hspace{10mm} A:=A(\sigma), \label{RA} \\
	&&{\cal R}_{A0}:= {\cal R}_{A}|_{y=0} \label{RA0},
\end{eqnarray}
$m_{\rm ADM}$ is the ADM mass, $m_{\rm ext}$
is the integral of the energy density outside $\sigma_0$,
\begin{equation}
	m_{\rm ext}:=2\pi\int^\infty_0dy{\cal R}_A^3\bar\rho,
	\label{mextdef}
\end{equation}
and the quantity ${\bar J_y}$ is defined as
\begin{equation}
	\left(8\pi\bar{J}_y\right)^2:=\frac{A^2}{6\pi}\int_{\sigma_y}v_av^adA,
\label{Jydef}
\end{equation}
which could be interpreted as the contribution of ``angular momentum'' on $\sigma_y$. 
This ``angular momentum'' $\bar J_y$ is defined in the following discussion.
If the space has the axial symmetry, the corresponding Killing vector $\phi^a$ exists, which runs from 0 to $2\pi$. 
Then, the Komar angular momentum $J$ can be introduced.
The Cauchy-Schwarz inequality gives the following inequality 
\footnote{
Equality in \eqref{Komar} is achieved if anf only if  $v^a$ is proportional to $\phi^a$, that is $v^a = C \phi^a$ with some constant $C$. However, this does not hold generically, not even in the Kerr solution. 
Therefore, the effect of the angular momentum is overestimated. 
This is one of the origins of the difficulty in proving the Penrose inequality with angular momentum~\cite{Lee2022b,Lee2022a}.  
}
\begin{equation}
\int_{\sigma_y} v^a v_a dA \int_{\sigma_y} \phi^a \phi_a dA \ge \left( \int_{\sigma_y} v^a \phi_a dA \right)^2 = (8\pi J)^2,
\label{Komar}
\end{equation}
where $J$ is the Komar angular momentum.
If the surface $\sigma_0$ is an exact sphere, we have 
\begin{equation}
\int_{\sigma_y} \phi^a \phi_a dA = \frac{A^2}{6 \pi},
\end{equation}
and the left-hand side of Eq.~\eqref{Komar} is written as 
\begin{equation}
\int_{\sigma_y} v^a v_a dA \int_{\sigma_y} \phi^a \phi_a dA = \frac{A^2}{6 \pi} \int_{\sigma_y} v^a v_a dA.
\end{equation}
Then, it may be natural to define ${\bar J_y}$ as Eq. \eqref{Jydef}.
For the last inequality in Eq.~\eqref{former.areaineq}, we have used the relation ${\cal R}_{A}={\cal R}_{A0}\exp{(y/2)}$ that holds in the IMCF, 
and have defined $\bar{J}_{\rm min}$ as $\bar{J}_{\rm min}:=\displaystyle \min_{\{\sigma_y\}}{\bar J_y}$.


\section{Attractive gravity probe surfaces}
\label{Secdef}

In Refs.~\cite{Shiromizu2017,Izumi2021,Izumi2022}, we presented and proved
an inequality for surfaces existing in weak gravity region,
which is a generalization of the Riemannian Penrose inequality. 
In Ref.~\cite{Lee2022b}, the further generalizations, called the longitudinal attractive gravity probe surfaces (LAGPSs), are discussed. 
Moreover, the proof of the generalized Riemannian Penrose inequality is refined by weakening the conditions imposed on the surface. 
The surface that satisfies the weakened conditions is called a refined attractive gravity probe surface.

On the other hand, a generalization of the photon sphere, called a dynamically transversely trapping surface (DTTS), was introduced in Ref.~\cite{Yoshino2019}, 
and an inequality similar to the generalized Riemannian Penrose inequality was derived. 
The concept of the DTTS was generalized in a similar way to the attractive gravity probe surface in Ref.~\cite{Lee2022b}, 
and was named a transverse attractive gravity probe surface (TAGPS). 
Here, we summarize their definitions.

\subsection{Longitudinal attractive gravity probe surface (LAGPS)}

In the generalization of the Riemannian Penrose inequality
an attractive gravity probe surface is introduced first as a surface existing in weak gravity region~\cite{Izumi2021}. 
The definition is as follows.
\begin{definition}
	\label{def..laspsk}
	\underline{Original AGPS (LAGPS-k)}: 
	{\it Suppose $\Sigma$ to be a smooth three-dimensional manifold with a positive definite metric $q_{ab}$. 
	A smooth compact surface $\sigma_0$ in $\Sigma$ is called an attractive gravity probe surface (AGPS) with a parameter $\alpha$ ($\alpha>-1/2$) if 
	\begin{eqnarray}
	&&k>0, \\ 
	&&r^a D_a k \ge \alpha k^2 \label{rDk>k2}
	\end{eqnarray}
	 are satisfied, where $r^a$ is the outward unit normal vector to $\sigma_0$, $k:=D_a r^a$  
	and $D_a$ is the covariant derivative with respect to the metric $q_{ab}$.}
\end{definition}
In this paper, we call this original AGPS an LAGPS-k to distinguish it from other surfaces. 
To define $r^a D_a k$, a foliation is required in the neighborhood of $\sigma_0$ on $\Sigma$. However, 
we do not specify it here because our proof works for any smooth foliation. 

Another version of the AGPS is introduced in Ref.~\cite{Lee2022b}. The definition is as follows.
\begin{definition}
	\label{def..laspsr}
	\underline{LAGPS-r}: 
	{\it In the definition for LAGPS-k (Def.~\ref{def..laspsk}), instead of Eq.~\eqref{rDk>k2}, 
	\begin{eqnarray}
	&&r^a D_a k\ge-(1-\gamma_L) {}^{(2)}R
	\label{defineqLAGPS-r}
	\end{eqnarray}
	is imposed, where $\gamma_L$ is a constant. Then, the surface is called an LAGPS-r.}
\end{definition}

The proof for the generalized Riemannian Penrose inequality has been improved~\cite{Izumi2022}, where the following refined definition of the AGPS is introduced.
\begin{definition}
	\label{def..refinedasps}
	\underline{Refined AGPS}: 
	{\it A refined AGPS is defined a surface such that, in the definition for LAGPS-k (Def.~\ref{def..laspsk}), instead of Eq.~\eqref{rDk>k2}, 
	\begin{eqnarray}
	&&{}^{(2)}R- 2k {\cal D}^2 \left(k^{-1}\right) \ge \left( 2\alpha + \frac{3}{2} \right) k^2
	\label{defineqrAGPS}
	\end{eqnarray}
	is imposed, where $\alpha$ ($\alpha >-1/2$) is a constant.}
\end{definition}
This definition is better than the other two in the sense that the foliation around $\sigma_0$ is not 
required, as the conditions for the refined AGPS are written only in
terms of the quantities that represents how $\sigma_0$ is embedded in $\Sigma$%
~\footnote{Note that Ref.~\cite{Izumi2022} gives a proof of the area inequality for the refined AGPS in higher dimensions. 
Since the Geroch monotonicity under the IMCF does not hold in higher dimensions, a different approach is required
  for the proof.}.

\subsection{Transverse attractive gravity probe surfaces (TAGPS)}

In Ref.~\cite{Yoshino2019}, a generalization of the photon sphere, called a dynamically transversely trapping surface (DTTS), is introduced, which is defined based on the geodesics for photons. 
Generalizations of a DTTS, which can exist in weak gravity region, are introduced in Ref.~\cite{Lee2022b}. 
The spirit of the definition is similar to that of the AGPS. We have the following two definitions of the generalizations.
\begin{definition}
	\label{def..taspsk}
	\underline{TAGPS (TAGPS-k, TAGPS-r)}: 
	{\it Suppose $S$ to be a smooth three-dimensional timelike hypersurface with a metric $p_{ab}$ in 4-dimensional spacetime $M$. 
	A smooth compact surface $\sigma_0$ in $S$ is called (i) a transverse attractive gravity probe surface -k (TAGPS-k) with a parameter $\beta$ ($\beta>-1/2$)
	and (ii) a transverse attractive gravity probe surface -r (TAGPS-r) with a parameter $\gamma_T$ if 
	\begin{eqnarray}
	&&(i)\quad \kappa =0, \quad \max(\bar{K}_{ab}k^ak^b)\le-\beta k \quad \mbox{and}\quad n^a \bar{D}_a \kappa\le 0,  \\
	&&(ii) \quad \kappa=0,  \quad \max(\bar{K}_{ab}k^ak^b)=0 \quad \mbox{and} \quad n^a \bar{D}_a \kappa\le {}^{(2)}R(1-\gamma_T)
	\end{eqnarray}
	are satisfied, respectively,
        where $k^a$ is an arbitrary null tangent vector to $S$ that satisfies
        $n^a={p^a}_bk^b$. 
	 Here, the variables in the above equations are defined in Sec.~\ref{Decomp}.}
\end{definition}


\section{Areal inequalities}
\label{lagps}

Areal inequalities, which are generalizations of the Riemannian Penrose inequality, have been proven in Refs.~\cite{Izumi2021,Izumi2022,Yoshino2019}
for the AGPSs. 
The proof based on the Geroch monotonicity is applicable to the other surfaces defined in the previous section~\cite{Lee2022b}. 
Keeping the non-negative contributions which are often removed in the derivation of inequalities, we obtain the refined versions of the inequalities. 
Then, the contributions of the angular momentum and electromagnetic charges become clear, and the extremality conditions for these surfaces can be obtained 
(see Sec.~\ref{cor}). 
In this section, we derive the refined inequalities.

\subsection{Area inequalities for longitudinal attractive gravity probe surface}

Suppose, in this subsection, that $\sigma_0$ is the LAGPS-k or LAGPS-r on a maximal slice $\Sigma$.
On a maximal surface $\Sigma$ satisfying $K_{(r)}=-\kappa $, Eq.~\eqref{liederiv.k} can be rewritten with Eq.~\eqref{em.energy} and the Einstein equation $G_{ab}=8\pi T_{ab}$ as
\begin{eqnarray}
	\label{liederiv.k2}
	r^aD_ak 
	&=&\frac{1}{2}{}^{(2)}R-(E_ar^a)^2-(B_ar^a)^2-\frac{3}{4}k^2-\frac{3}{4}\kappa^2-8\pi (\rho^{\rm(em)}_\parallel+\rho^{(m)}+\rho_{\rm gw})-v_av^a-\varphi^{-1}{\cal D}^2\varphi
	\nonumber \\
	&\le& \frac{1}{2}{}^{(2)}R-(E_ar^a)^2-(B_ar^a)^2-\frac{3}{4}k^2-8\pi (\rho^{\rm(em)}_\parallel+\rho^{(m)}+\rho_{\rm gw})-v_av^a-\varphi^{-1}{\cal D}^2\varphi.
\end{eqnarray}
Integrating Eq.~(\ref{liederiv.k2}) over $\sigma_0$, which is the LAGPS-k or LAGPS-r, we have the upper bound of the integral of $k^2$,
\begin{align}
	\label{liederiv.lagpsk}
	&\left(1+\frac{4}{3}\alpha\right)\int_{\sigma_0}k^2dA
	\nonumber\\ 
	& \qquad 
	\le\frac{16\pi}{3}-\frac{2}{3}\int_{\sigma_0}dA\Big[2(E_ar^a)^2+2(B_ar^a)^2+16\pi(\rho^{\rm(em)}_\parallel+\rho^{(m)}+\rho_{\rm gw})
+2v_av^a\Big]\nonumber\\ 
	& \qquad 
	\le\frac{16\pi}{3}-\frac{16\pi}{3}\frac{q^2}{{\cal R}_{A0}^2}-\frac{32\pi}{3}\Phi_0-\frac{128\pi^2}{3}{\cal R}_{A0}^2\left(\bar{\rho}^{(0){\rm (m)}}+\bar{\rho}^{(0)}_{\rm gw}\right)-32\pi\frac{\bar J_0^2}{{\cal R}_{A0}^4},
\end{align}
or
\begin{align}
	\label{liederiv.lagpsr}
	\int_{\sigma_0}k^2dA &\le
	\frac{16\pi}{3}\left(3-2\gamma_L\right)-\frac{2}{3}\int_{\sigma_0}dA\Big[2(E_ar^a)^2+2(B_ar^a)^2+16\pi(\rho^{\rm(em)}_\parallel+\rho^{(m)}+\rho_{\rm gw})+2v_av^a\Big]\nonumber\\ 
	& \le \frac{16\pi}{3}\left(3-2\gamma_L\right)-\frac{16\pi}{3}\frac{q^2}{{\cal R}_{A0}^2}-\frac{32\pi}{3}\Phi_0-\frac{128\pi^2}{3}{\cal R}_{A0}^2\left(\bar{\rho}^{(0){\rm (m)}}+\bar{\rho}^{(0)}_{\rm gw}\right)-32\pi\frac{\bar J_0^2}{{\cal R}_{A0}^4},
\end{align}
respectively, where 
$\bar{\rho}^{(0){\rm (m)}}$ and $\bar{\rho}^{(0)}_{\rm gw}$ are the area-averaged energy densities of matter fields and of the gravitational field on $\sigma_0$ defined as
\begin{align}
	\bar{\rho}^{(0){\rm (m)}} :=\frac{1}{A_0}\int_{\sigma_0} \rho^{\rm (m)} dA, \hspace{10mm}
	\bar{\rho}^{(0)}_{\rm gw} :=\frac{1}{A_0}\int_{\sigma_0} \rho_{\rm gw} dA,
\end{align}
and $\Phi_0$ is the areal integral of contribution of the tangent components of electric and magnetic fields to the energy density on $\sigma_0$,
\begin{equation}
	\label{EMparallel}
	\Phi_0:=\int_{\sigma_0}\rho^{\rm(em)}_\parallel dA.
\end{equation}

Combining Eq.~\eqref{former.areaineq} with Eq.~\eqref{liederiv.lagpsk}, we obtain the following theorem about the relation between the ADM mass and the area of the LAGPS-k:


\begin{theorem}
	\label{thm..laspsk}
{\it	Let $\Sigma$ be an asymptotically flat spacelike maximal hypersurface foliated by the inverse mean curvature flow $\lbrace \sigma_y \rbrace_{y \in {\mathbb R}}$ with $\sigma_y \approx S^2$, where $\sigma_0$ is an LAGPS-k.
	Assuming that 
	no electromagnetic charge exists outside $\sigma_0$, we have an inequality about the areal radius ${\cal R}_{A0}$ of the LAGPS-k $\sigma_0$,
	\begin{align}
		\label{clagps-k}
		&m_{\rm ADM}-\left(m_{\rm ext}+\frac{3}{3+4\alpha}m_{\rm int}\right)\nonumber\\
		&\qquad\ge\frac{1+2\alpha+\Phi_0}{3+4\alpha}{\cal R}_{A0}+\frac{2(1+\alpha)}{3+4\alpha}\frac{q^2}{{\cal R}_{A0}}+\frac{1}{{\cal R}_{A0}^3}\left(\bar J^2_{\rm min}+\frac{3}{3+4\alpha}\bar J_0^2\right)\nonumber\\
		&\qquad\ge\frac{1+2\alpha+\Phi_0}{3+4\alpha}{\cal R}_{A0}+\frac{2(1+\alpha)}{3+4\alpha}\frac{q^2}{{\cal R}_{A0}}+\frac{2(3+2\alpha)}{3+4\alpha}\frac{\bar J_{\rm min}^2}{{\cal R}_{A0}^3},
	\end{align}
	where $m_{\rm ext}$ and $m_{\rm int}$ are defined by Eq.~\eqref{mextdef} and 
	\begin{equation}
		m_{\rm int}:=\frac{4}{3}\pi{\cal R}_{A0}^3\left(\bar{\rho}^{(0){\rm (m)}}+\bar{\rho}^{(0)}_{\rm gw}\right),
	\end{equation}
	respectively.}
\end{theorem}

For $\alpha\to\infty$ ($\sigma_0$ to be a minimal surface) and $\alpha=0$ ($\sigma_0$ to be an LTS), we have
\begin{equation}
	\label{cms.area.ineq}
	m_{\rm ADM}-m_{\rm ext}\ge\frac{{\cal R}_{A0}}{2}+\frac{1}{2}\frac{q^2}{{\cal R}_{A0}}+\frac{\bar J^2_{\rm min}}{{\cal R}_{A0}^3}
\end{equation}
and
\begin{equation}
	\label{clts.area.ineq}
	m_{\rm ADM}-\left(m_{\rm ext}+m_{\rm int}\right)\ge\frac{{\cal R}_{A0}}{3}(1+\Phi_0)+\frac{2}{3}\frac{q^2}{{\cal R}_{A0}}+2\frac{\bar J^2_{\rm min}}{{\cal R}_{A0}^3},
\end{equation}
respectively.

Combination of Eqs.~\eqref{former.areaineq} and \eqref{liederiv.lagpsr} gives the following theorem about the areal radius 
of the LAGPS-r:


\begin{theorem}
	\label{thm..laspsr}
	{\it Let $\Sigma$ be an asymptotically flat spacelike maximal hypersurface foliated by the inverse mean curvature flow $\lbrace \sigma_y \rbrace_{y \in {\mathbb R}}$ with $\sigma_y \approx S^2$, where $\sigma_0$ is an LAGPS-r.
	Assuming that 
	no electromagnetic charge exists outside the LAGPS-r $\sigma_0$, we have an inequality for the LAGPS-r $\sigma_0$,
	\begin{align}
		\label{clagps-r}
		m_{\rm ADM}-\left(m_{\rm ext}+m_{\rm int}\right)&\ge\frac{1}{3}\left(\gamma_L+\Phi_0\right){\cal R}_{A0}+\frac{2}{3}\frac{q^2}{{\cal R}_{A0}}+\frac{1}{{\cal R}_{A0}^3}\left(\bar J^2_{\rm min}+\bar J_0^2\right)\nonumber\\
		&\ge\frac{1}{3}\left(\gamma_L+\Phi_0\right){\cal R}_{A0}+\frac{2}{3}\frac{q^2}{{\cal R}_{A0}}+2\frac{\bar J^2_{\rm min}}{{\cal R}_{A0}^3}.
	\end{align}
	}
\end{theorem}

For $\gamma_L=1$ ($\sigma_0$ to be an LTS), we have the 
same inequality as Eq.~\eqref{clts.area.ineq}.


\subsection{Areal inequalities for transverse attractive gravity probe surface}
\label{tagps}

In this subsection, we show the areal inequalities for transverse attractive gravity probe surfaces (TAGPSs), which are defined based on the variation of geometric quantities with respect to the ``transverse" direction to the surface~\cite{Lee2022b}. 
Hence, we suppose, in this subsection, that $\sigma_0$ is a TAGPS-k or TAGPS-r.   
For the derivation, we introduce a useful inequality for the quantity $\max(\bar{K}_{ab}k^ak^b)$ (see more details in  Refs.~\cite{Lee2022b, Lee2020}), 
\begin{align}
	\label{useful.ineq}
	\max(\bar{K}_{ab}k^ak^b)\ge\bar{K}_{(n)}+\frac{1}{2}k.
\end{align}
In the similar arguments to the LAGPS case, through the Einstein equation $G_{ab}=8\pi T_{ab}$, the quantity
$G_{ab}r^ar^b$ in Eq.~\eqref{liederiv.kappa} can be replaced by the radial pressure, that is, 
\begin{align}
	\label{liederiv.kappa2}
	n^a\bar{D}_a\kappa=&-\frac{1}{2}{}^{(2)}R+(E_ar^a)^2+(B_ar^a)^2+\frac{1}{4}k^2-k\bar{K}_{(n)}-\frac{3}{4}\kappa^2
\nonumber\\
	&\quad
	-8\pi\left(\rho_\parallel^{({\rm em})}+P_r^{(m)}+P_r^{({\rm gw})}\right)
	+v_av^a+N^{-1}{\cal D}^2N,
\end{align}
where we have used Eq.~\eqref{em.energy}.
Here, $P_r^{(m)}$ is the radial pressure of the matter fields, that is, $P_r^{(m)}:=T_{ab}^{(m)}r^ar^b$, and $P_r^{({\rm gw})}$
is  the radial pressure of the gravitational field defined by
\begin{equation}
	8\pi P_r^{({\rm gw})} :=\frac{1}{2}(\tilde \kappa_{ab} \tilde \kappa^{ab}+\tilde k_{ab}\tilde k^{ab})=8\pi \rho_{{\rm gw}}.
\end{equation}
In the following, we assume the existence of the IMCF 
$\{\sigma_y\}$ with $y\ge 0$ along which the non-negativity
of $k=1/\varphi$ is guaranteed. 
Taking the integral of Eq.~\eqref{liederiv.kappa2} over $\sigma_0$ and using Eq.~\eqref{useful.ineq}, we have an inequality for the integral of $k^2$ over $\sigma_0$
\begin{align}
	\label{liederiv.tagpsk}
	&\left(1+\frac{4}{3}\beta\right)\int_{\sigma_0}k^2dA
	\nonumber\\ & \qquad \le 
	\frac{16\pi}{3}-\frac{2}{3}\int_{\sigma_0}dA\Big[2(E_ar^a)^2+2(B_ar^a)^2-16\pi(\rho^{\rm(em)}_\parallel +P_r^{(m)}+P_r^{({\rm gw})})+2v_av^a\Big]
	\nonumber\\ & \qquad \le 
	\frac{16\pi}{3}-\frac{16\pi}{3}\frac{q^2}{{\cal R}_{A0}^2}+\frac{32\pi}{3}\Phi_0+\frac{128\pi^2}{3}{\cal R}_{A0}^2\left(\bar{P}_r^{(0){\rm (m)}}+\bar{P}_r^{(0)({\rm gw})}\right)-32\pi\frac{\bar J_0^2}{{\cal R}_{A0}^4},
\end{align}
for the TAGPS-k (where $k>0$ has been used), or
\begin{align}
	\label{liederiv.tagpsr}
	\int_{\sigma_0}k^2dA\le&\frac{16\pi}{3}\left(3-2\gamma_T\right)-\frac{2}{3}\int_{\sigma_0}dA\Big[2(E_ar^a)^2+2(B_ar^a)^2-16\pi(\rho^{\rm(em)}_\parallel+P_r^{(m)}+P_r^{({\rm gw})})+2v_av^a\Big]\nonumber\\
	\le&\frac{16\pi}{3}\left(3-2\gamma_T\right)-\frac{16\pi}{3}\frac{q^2}{{\cal R}_{A0}^2}+\frac{32\pi}{3}\Phi_0+\frac{128\pi^2}{3}{\cal R}_{A0}^2\left(\bar{P}_r^{(0){\rm (m)}}+\bar{P}_r^{(0)({\rm gw})}\right)-32\pi\frac{\bar J_0^2}{{\cal R}_{A0}^4},
\end{align}
for the TAGPS-r, where $\bar{P}_r^{(0){\rm (m)}}$ and $\bar{P}_r^{(0)({\rm gw})}$ are the area-averaged radial pressure on $\sigma_0$ defined as
\begin{align}
	\bar{P}_r^{(0){\rm (m)}}&:=\frac{1}{A_0}\int_{\sigma_0} P_r^{(m)}dA,\\
	\bar{P}_r^{(0)({\rm gw})}&:=\frac{1}{A_0}\int_{\sigma_0}P_r^{({\rm gw})}dA.
\end{align}

Using Eqs.~\eqref{former.areaineq} and \eqref{liederiv.tagpsk}, one can obtain the following theorem for the area of the TAGPS-k:


\begin{theorem}
	\label{thm..taspsk}
{\it	Let $\Sigma$ be an asymptotically flat spacelike maximal hypersurface foliated by the inverse mean curvature flow $\lbrace \sigma_y \rbrace_{y \in {\mathbb R}}$ with $\sigma_y \approx S^2$, where $\sigma_0$ is a TAGPS-k.
	Assuming that 
	no electromagnetic charge exists outside the TAGPS-k $\sigma_0$, we have an inequality for the TAGPS-k $\sigma_0$,
	\begin{align}
		\label{ctagps-k}
		&m_{\rm ADM}-\left(m_{\rm ext}-\frac{3}{3+4\beta}p_r^{({\rm int})}\right)\nonumber\\
		&\qquad\ge\frac{1+2\beta-\Phi_0}{3+4\beta}{\cal R}_{A0}+\frac{2(1+\beta)}{3+4\beta}\frac{q^2}{{\cal R}_{A0}}+\frac{1}{{\cal R}_{A0}^3}\left(\bar J^2_{\rm min}+\frac{3}{3+4\beta}\bar J_0^2\right)\nonumber\\
		&\qquad\ge\frac{1+2\beta-\Phi_0}{3+4\beta}{\cal R}_{A0}+\frac{2(1+\beta)}{3+4\beta}\frac{q^2}{{\cal R}_{A0}}+\frac{2(3+2\beta)}{3+4\beta}\frac{\bar J_{\rm min}^2}{{\cal R}_{A0}^3},
	\end{align}
	where $p_r^{({\rm int})}$ is defined as
	\begin{equation}
		p_r^{({\rm int})}:=\frac{4}{3}\pi{\cal R}_{A0}^3\left(\bar{P}_r^{(0){\rm (m)}}+\bar{P}_r^{(0)({\rm gw})}\right).
	\end{equation}
}
\end{theorem}

For $\beta\to\infty$ ($\sigma_0$ to be a minimal surface) and $\beta=0$ ($\sigma_0$ to be a DTTS), we have
\begin{equation}
	m_{\rm ADM}-m_{\rm ext}\ge\frac{{\cal R}_{A0}}{2}+\frac{1}{2}\frac{q^2}{{\cal R}_{A0}}+\frac{\bar J^2_{\rm min}}{{\cal R}_{A0}^3}
\end{equation}
and
\begin{equation}
	\label{cdtts.area.ineq}
	m_{\rm ADM}-\left(m_{\rm ext}-p_r^{({\rm int})}\right)\ge\frac{{\cal R}_{A0}}{3}(1-\Phi_0)+\frac{2}{3}\frac{q^2}{{\cal R}_{A0}}+2\frac{\bar J^2_{\rm min}}{{\cal R}_{A0}^3},
\end{equation}
respectively.

Similarly, Eqs.~\eqref{former.areaineq} and \eqref{liederiv.tagpsr} give the following theorem for the area of the TAGPS-r:


\begin{theorem}
	\label{thm..tagpsr}
{\it	Let $\Sigma$ be an asymptotically flat spacelike maximal hypersurface foliated by the inverse mean curvature flow $\lbrace \sigma_y \rbrace_{y \in {\mathbb R}}$ with $\sigma_y \approx S^2$, where $\sigma_0$ is a TAGPS-r.
	Assuming that 
	no electromagnetic  charge exists outside the TAGPS-r $\sigma_0$, we have an inequality for the TAGPS-r $\sigma_0$,
	\begin{align}
		\label{ctagps-r}
m_{\rm ADM}-\left(m_{\rm ext}-p_r^{({\rm int})}\right)&\ge\frac{1}{3}\left(\gamma_T-\Phi_0\right){\cal R}_{A0}+\frac{2}{3}\frac{q^2}{{\cal R}_{A0}}+\frac{1}{{\cal R}_{A0}^3}\left(\bar J^2_{\rm min}+\bar J_0^2\right)\nonumber\\
		&\ge\frac{1}{3}\left(\gamma_T-\Phi_0\right){\cal R}_{A0}+\frac{2}{3}\frac{q^2}{{\cal R}_{A0}}+2\frac{\bar J^2_{\rm min}}{{\cal R}_{A0}^3}.
	\end{align}
}
\end{theorem}

For $\gamma_T=1$ ($\sigma_0$ to be a DTTS), we have the same inequality as Eq.~\eqref{cdtts.area.ineq}.


\subsection{Areal inequalities for refined attractive gravity probe surface}
\label{ragps}

We have the following inequality for the refined AGPS $\sigma_0$,
\begin{align}
	\left(1+\frac{4}{3}\alpha\right)\int_{\sigma_0}dAk^2\le\frac{4}{3}\int_{\sigma_0}dA\left(-k{\cal D}^2\left(k^{-1}\right)+\frac{1}{2}{}^{(2)}R\right)\le\frac{16\pi}{3}.
	\label{refinedAGPSonsigma}
\end{align}
Then we obtain the following theorem from Eq.~\eqref{former.areaineq}:
%
%
\begin{theorem}
	\label{thm..refined-agps}
{\it	Let $\Sigma$ be an asymptotically flat spacelike maximal hypersurface foliated by the inverse mean curvature flow $\lbrace \sigma_y \rbrace_{y \in {\mathbb R}}$ with $\sigma_y \approx S^2$, where $\sigma_0$ is a refined AGPS.
	Assuming that 
	no electromagnetic charge exists outside $\sigma_0$, we have an inequality for the refiend AGPS $\sigma_0$,
\begin{equation}
	\label{cragps}
	m_{\rm ADM}-m_{\rm ext}\ge\frac{1+2\alpha}{3+4\alpha}{\cal R}_{A0}+\frac{1}{2}\frac{q^2}{{\cal R}_{A0}}+\frac{\bar{J}_{\rm min}^2}{{\cal R}^3_{A0}}.
\end{equation}
}
\end{theorem}
Note that Eq.~\eqref{cragps} does not include the integral on $\sigma_0$, such as $m_{\rm int}$, and thus the inequality is expected to be saturated. 
This may mean that the definition of a refined AGPS is better than the others.
For $\alpha\to\infty$ ($\sigma_0$ to be a minimal surface), we have
the same inequality as Eq.~\eqref{cms.area.ineq}.


\section{Conditions for Saturation and Extremality}
\label{cor}

It is interesting to consider the contributions only of the electric and magnetic charges. 
In the right-hand sides of Eqs.~\eqref{clagps-k}, \eqref{clagps-r}, \eqref{ctagps-k}, \eqref{ctagps-r} and \eqref{cragps}, 
${\bar J}_{\rm min}^2$ terms give non-negative contributions, and thus the inequalities are satisfied even if ${\bar J}_{\rm min}^2$ 
terms are removed. Then, the inequalities show the lower bound of mass under the given electromagnetic charge, which can be 
interpreted as the extremality conditions for the AGPSs, 
similar to the horizon for the Reissner-Nordstr\"{o}m black holes.
In this way, we are led to the concept of extremality
for surfaces other than the horizons. 
The extremality conditions can be simplified by taking the arithmetic-geometric mean for our area inequalities of Eqs.~\eqref{clagps-k}, \eqref{clagps-r}, \eqref{ctagps-k} and \eqref{ctagps-r}. 
With respect to the extremality conditions for the angular momentum, see also Corollary 2 and 4 in Ref.~\cite{Lee2022a} or Eq.~(110) in Ref.~\cite{Lee2022b}. 

Before discussing the extremality condition, we investigate the condition of saturation for the areal inequality. 
In the original inequalities for AGPSs presented in Refs.~\cite{Shiromizu2017,Izumi2021,Izumi2022}, where contributions 
from the energy density of matters, the Maxwell field, gravitational waves and angular momentum are not included, 
the equality holds if and only if three-dimensional manifold $\Sigma$ is isometric to the time-symmetric slice of the Schwarzschild solution. 
In this case, the AGPS becomes the $r$-constant surface where the equality holds in the inequality with $\alpha$ in the definition. 
Including the electromagnetic contributions, we have a similar result, that is,  the equality holds if and only if three-dimensional manifold $\Sigma$ is isometric to the time-symmetric slice of the Reissner-Nordstr\"{o}m solution\footnote{
Note that the connectedness of the AGPS is assumed here, because our proof relies on the IMCF. 
}.
Suppose the energy condition holds, which means that the energy density of matter $\rho^{(m)}$ is non-negative.
In the areal inequalities for the LAGPSs and the refined AGPS, given by Eqs.~\eqref{clagps-k}, \eqref{clagps-r} and \eqref{cragps}, 
even if $m_{\rm ext}$, $m_{\rm int}$, $\Phi_0$ and $\bar J^2_{\rm min}$ are removed, the inequalities still hold due to the non-negativity of 
these variables, that is,
\begin{eqnarray}
&&m_{\rm ADM}\ge\frac{1+2\alpha}{3+4\alpha}{\cal R}_{A0}+\frac{2+2\alpha}{3+4\alpha}\frac{q^2}{{\cal R}_{A0}}, \label{maxineq1} \\
&&m_{\rm ADM}\ge\frac{1}{3}\gamma_L{\cal R}_{A0}+\frac{2}{3}\frac{q^2}{{\cal R}_{A0}},\label{maxineq2}
\end{eqnarray}
and
\begin{eqnarray}
&&m_{\rm ADM}\ge\frac{1+2\alpha}{3+4\alpha}{\cal R}_{A0}+\frac{1}{2}\frac{q^2}{{\cal R}_{A0}} \label{maxineq3}
\end{eqnarray}
hold, respectively. 
The equalities hold if and only if all the non-negative contributions which we remove to derive the inequalities vanish, 
and the equality holds in Eqs.~\eqref{rDk>k2},  \eqref{defineqLAGPS-r} or \eqref{defineqrAGPS}.
This implies that $\Sigma$ enjoys the spherical symmetry%
\footnote{For detail, see Ref.~\cite{Shiromizu2017}.} 
and is the Einstein-Maxwell vacuum solution, that is, the time-symmetric slice of the Reissner-Nordstr\"{o}m solution%
\footnote{The Reissner-Nordstr\"{o}m solution can be super-extremal.} 
due to the Birkhoff theorem. Consequently, the AGPS is the $r$-constant surface where the equality holds in the inequality 
with $\alpha$ in the definition.

The original AGPS (LAGPS-k) is proposed as an extension of the LTS, which is a generalization of the photon sphere.
This means that the LTS is a special case of the LAGPS-k with $\alpha=0$ and that the LTS in the case where equality occurs in the areal inequality coincides with the photon sphere. 
This is true for the LTS in the Einstein-Maxwell system, that is, 
if equalities hold in the areal inequalities \eqref{maxineq1}, \eqref{maxineq2} or \eqref{maxineq3}, the LTS ({\it i.e.} the LAGPS with $\alpha=0$) is the photon sphere in the Reissner-Nordstr\"{o}m solution. 
Note that in the case where equality holds in the areal inequality, the photon sphere does not coincide with the refined AGPS with $\alpha =0$. 
This is obvious from the definition; the condition \eqref{defineqrAGPS} is different from Eq.~\eqref{rDk>k2}, if ${}^{(3)}R \neq 0$. 
The refined AGPS coincide with the photon sphere for
\begin{equation}
\alpha = \frac{-3m^2+2q^2+m\sqrt{9m^2-8q^2}}{8\left(m^2-q^2\right)}.
\end{equation}

Let us move on to the discussion of the extremality condition for AGPSs.
For the LAGPS-k and TAGPS-k, from Eqs.~\eqref{clagps-k} and \eqref{ctagps-k}, removing the ${\bar J}_{\rm min}^2$ terms and taking the arithmetic-geometric mean, we have
\begin{equation}
	\label{clagps-k.cor}
	\Delta m_\pm^{\rm k}\ge2|q|\sqrt{\frac{2(1+\alpha_\pm)(1+2\alpha_\pm\pm\Phi_0)}{(3+4\alpha_\pm)^2}},
\end{equation}
where
\begin{align}
	\Delta m_+^{\rm k}&=m_{\rm ADM}-\left(m_{\rm ext}+\frac{3}{3+4\alpha_+}m_{\rm int}\right),\\
	\Delta m_-^{\rm k}&=m_{\rm ADM}-\left(m_{\rm ext}-\frac{3}{3+4\alpha_-}p_r^{({\rm int})}\right),
\end{align}
and we introduced $\alpha_+=\alpha$ and $\alpha_-=\beta$ to express the inequalities in a unified manner. 
Note that, for the TAGPS-k, $\alpha_-\ge(\Phi_0-1)/2$ is required to obtain Eq.~\eqref{clagps-k.cor}.
For $\alpha_\pm=0$, where the surface becomes an LTS or a DTTS, Eq.~\eqref{clagps-k.cor} gives
a consistent result with Eqs.~(22) and (37) in Ref.~\cite{Lee2020}.
We show that Eq.~\eqref{clagps-k.cor} leads to
the concept of extremality for the LAGPS-k ({\it i.e.} for ``+'' case).
Since the ADM mass $m_{\rm ADM}$  satisfies $m_{\rm ADM}\ge\Delta m_+^{\rm k}$, 
Eq.~\eqref{clagps-k.cor} can be reduced to
\begin{equation}
  m_{\rm ADM}\ge m_{\rm ADM}-m_{\rm ext}\ge2|q|\sqrt{\frac{2(1+\alpha)(1+2\alpha)}{(3+4\alpha)^2}},
  \label{Extremality-LAGPSk}
\end{equation}
where we have used the fact that $\Phi_0 \ge 0$ and $\alpha(>-1/2)>-3/4$. 
In the limit $\alpha\to\infty$ where the surface is reduced to a minimal surface, Eq.~\eqref{clagps-k.cor} gives
\begin{equation}
	m_{\rm ADM}\ge|q|,
\end{equation}
where the equality gives the extremality condition for the minimal surface,
as proved in Ref.~\cite{Gibbons1982} (see also Ref.~\cite{Khuri2013}).
Therefore, the equality of 
Eq.~\eqref{Extremality-LAGPSk} can be interpreted as the extremality condition on the charge for the LAGPS-k.

It would be helpful to interpret the meaning of the obtained
concept of extremality in the case of the Reissner-Nordstr\"om spacetime.
For a surface $r=\mathrm{constant}$
in the $t=\mathrm{constant}$ hypersurface in a Reissner-Nordstr\"om spacetime, $m_{\rm ext}=\Phi_0=0$ holds
because 
$\rho^{\rm (em)}_\parallel$ defined in Eq.~\eqref{Def:rho-em-parallel}
is zero due to spherical symmetry.
Then, Eq.~\eqref{clagps-k.cor} is reduced to
\begin{equation}
\frac{|q|}{m_{\rm ADM}} \,\le\,  \frac{3+4\alpha}{2\sqrt{2(1+\alpha)(1+2\alpha)}},
\end{equation}
where the right-hand side is a monotonically decreasing function
of $\alpha$ and becomes infinitely large in the limit $\alpha\to -1/2$. 
This inequality represents the condition that an LAGPS-k
with the parameter $\alpha$ can exist in the Reissner-Nordstr\"om spacetime.
To see this, let us consider the process of increasing the value of
${|q|}/m_{\rm ADM}$ from unity in the solution space.
As is well-known, the super-extremal Reissner-Nordstr\"om spacetime 
includes the anti-gravity region around the center. As ${|q|}/m_{\rm ADM}$ increases, the anti-gravity region becomes larger 
while the attractive gravity region decreases.
As a result, the maximum value of $\alpha$ in the collection 
of the AGPSs decreases in this process (see also Fig.~1 of Ref.~\cite{Lee2022b}).
Hence, for a given $\alpha$, there is a maximum value of $|q|/m_{\rm ADM}$
such that the AGPS with the parameter $\alpha$ can exist. The concept of extremality gives the threshold for the (non-)existence of 
the AGPS with the parameter $\alpha$.

For the LAGPS-r and TAGPS-r,  the arithmetic-geometric means of Eqs.~\eqref{clagps-r} and \eqref{ctagps-r} give
\begin{equation}
	\label{clagps-r.cor}
	\Delta m_\pm^{\rm r}\ge\frac{2\sqrt{2}}{3}|q|\sqrt{\gamma_\pm\pm\Phi_0}, 
\end{equation}
where
\begin{align}
	\Delta m_+^{\rm r}&=m_{\rm ADM}-\left(m_{\rm ext}+m_{\rm int}\right),\\
	\Delta m_-^{\rm r}&=m_{\rm ADM}-\left(m_{\rm ext}-p_r^{({\rm int})}\right),
\end{align}
and we have introduced $\gamma_+=\gamma_L$ and $\gamma_-=\gamma_T$ to express the inequalities in a unified manner. 
Here, for the TAGPS-r, the condition $\gamma_T\ge\Phi_0$ is required.
Note that $\Delta m_\pm^{\rm r}$ are independent of $\gamma_\pm$.
In the case $\gamma_\pm=1$ where the surface becomes an LTS or a DTTS, Eq.~\eqref{clagps-r.cor} again gives 
a consistent result with those in Ref.~\cite{Lee2020}. 
The inequality of Eq.~\eqref{clagps-r.cor} for the LAGPS-r ({\it i.e.} the ``$+$'' case) can be simplified as 
\begin{equation}
	m_{\rm ADM}\ge\frac{2\sqrt{2}}{3}|q|\sqrt{\gamma_L}, 
\end{equation}
the equality of which may be interpreted as the extremality condition, but the relation of the parameter $\gamma_L$ to the minimal surface is nontrivial.

We can also derive the extremality condition for the refined AGPS. 
By taking the arithmetic-geometric mean of Eq.~\eqref{cragps} with ignoring $\bar{J}$ contribution, we have the extremality condition for the electromagnetic charges,
i.e., when the equalities hold in the following inequalities:
\begin{equation}
	m_{\rm ADM}\ge \, m_{\rm ADM}-m_{\rm ext}\ge|q|\sqrt{\frac{2+4\alpha}{3+4\alpha}}.  \label{extrAPGS}
\end{equation}
We also have the extremality condition for the angular momentum,
i.e., when the equalities hold in the following inequalities:
\begin{equation}
	m_{\rm ADM}\ge m_{\rm ADM}-m_{\rm ext}\ge \frac{1+2\alpha}{3+4\alpha}{\cal R}_{A0} + \frac{\bar{J}_{\rm min}^2}{{\cal R}_{A0}^3}
	\ge 2\frac{|\bar{J}_{\rm min}|}{{\cal R}_{A0}}\sqrt{\frac{1+2\alpha}{3+4\alpha}}.
\end{equation}

We give comments on the rigidity, the extremality condition and the positive energy theorem with charges\footnote{
We would like to thank D. Yoshida for making us realize the importance of this relation.
}.
It is known~\cite{Disconzi:2012es,Gibbons1982,Jang:1979zz,Khuri2013,Gibbons:1982fy,Weinstein2004} that the extremality condition $m_{\rm ADM}\ge |q|$ 
for minimal surfaces holds if and only if $\Sigma$ is the time-symmetric slice of the Majumdar-Papapetrou solution.
This is interpreted as the positive energy theorem with charges, which states $m_{\rm ADM}\ge 0$ and that the equality holds if and only if $\Sigma$ is isometric to the Euclid space. 
This rigidity condition for a single charged minimal surface implies that the extremality condition holds if and only if $\Sigma$ is the time-symmetric slice of the Reissner-Nordstr\"om spacetime.
Our extremality condition gives a similar statement. 
Let us consider the refined AGPS (the same is true for LAGPSs).  
Suppose that the boundary of $\Sigma$ is composed of one asymptotically flat end and a connected refined AGPS\footnote{
We expect that the statement is true even if multiple AGPSs exist. 
However, since in this paper the proof relies on the IMCF, the statement is guaranteed to hold only in the case with a single AGPS. 
The areal inequality for multiple AGPSs without the contributions of the Maxwell field has been proved~\cite{Izumi2021,Izumi2022}. 
A similar proof may work, and if so, the statement here would be extended to the case with multiple AGPSs. 
} with the fixed parameter $\alpha$.
Then, what we have shown is that the extremality condition~\eqref{extrAPGS} is satisfied, and that the rigidity holds, that is, the equality holds if and only if 
$\Sigma$ is isometric to a region outside the $r$-constant surface where the equality holds in inequality~\eqref{defineqrAGPS}, in the time-symmetric slice of the (super-extremal) Reissner-Nordstr\"om spacetime. 
The singularity of the super-extremal Reissner-Nordstr\"om spacetime
is outside of $\Sigma$ because it 
is hidden in the refined AGPS.

We also note that the extremality condition for the AGPSs
is given in terms of the global quantities above.
For the horizon, the extremality condition 
can be given by a local quantity as well, that is, the surface gravity
to be zero. Although it may be possible to introduce
the local definition of the extremality of the AGPSs,
this problem is postponed as a future issue.


\section{Inequality for asymptotically AdS spacetime}
\label{AdS}

It would be useful to derive the inequalities for asymptotically locally AdS spacetimes for application 
to the AdS/CFT correspondence~\cite{Maldacena:1997re}. 
Since the extremality conditions of charged black holes are discussed in the context of the swampland program~\cite{Wald:1974hkz,Hod:2002pm,Vafa:2005ui,Arkani-Hamed:2006emk,Sorce:2017dst,Izumi:2024rge}, 
those for AGPS are expected to be useful for the understanding of quantum gravity. 
Although we have a few definitions of AGPSs as shown in Sec.~\ref{Secdef}, here we only show the derivation of the inequality and the extremality condition for a refined AGPS.
We will comment on the differences from the other cases if there are. 

The system we consider here is the same as before but has the contribution of the negative cosmological constant which makes the asymptotically locally AdS structure.
Suppose that the metric of the spacetime $M$ considered here asymptotically approaches   to
\begin{eqnarray}
ds^2 = - f(r) dt^2 + f^{-1}(r)dr^2 +r^2 d \omega_\iota^2,
\label{metaAdS}
\end{eqnarray}
with
\begin{eqnarray}
f(r):= \iota - \frac{2m}{r}-\frac{\Lambda}{3}r^2,
\label{fiota}
\end{eqnarray}
where $\iota=\pm1$ or $0$, and $\Lambda$ is the negative cosmological constant. 
$d\omega_\iota^2$ is the metric of a two-dimensional compact and maximally symmetric surface
(for $\iota=0$, its topology is torus; for $\iota=-1$, multi-torus; for $\iota=1$, sphere). 
If the induced metric of
a spacelike hypersurface $\Sigma$ in $M$ asymptotically approaches
\begin{eqnarray}
dl^2 = f^{-1}(r)dr^2 +r^2 d \omega_\iota^2,
\end{eqnarray}
we call $\Sigma$ an asymptotically locally AdS space.
The decomposition of the spacetime can be done in the same way as that in Sec.~\ref{Decomp}. 
The Einstein tensor is replaced by the energy-momentum tensor though the Einstein equation, and here, the negative cosmological constant also contributes.
The Geroch energy should be modified to include the contribution of the negative cosmological constant.
The extended version of the Geroch energy has already been introduced in Refs.~\cite{Gibbons:1998zr,Boucher:1983cv,Lee:2015xha,Fischetti:2016fbh}. 
Here, we use the further extended version of the Geroch energy defined in Ref.~\cite{Izumi2021} which depends on the topology of the boundary, 
\begin{eqnarray}
	E(y):=\frac{A^{1/2}(\sigma_y)}{\left(4 \omega_\iota\right)^{3/2}}\int_{\sigma_y}\left(2{}^{(2)}R-k^2-\frac{4\Lambda}{3}\right)dA, 
\end{eqnarray}
where $\omega_\iota$ is the area of the compact two-dimensional surface with the metric $d\omega_\iota$.
If we take the limit where $\sigma_y$ goes to the spatial infinity,
this extended Geroch energy becomes the rescaled 
Abbott-Deser mass
$m=m_{\rm AD}$~\cite{Abbott:1981ff}
that is related to the Abbott-Deser mass $\mu_{\rm AD}$ as $m_{\rm AD}=(4\pi/\omega_\iota)\mu_{\rm AD}$%
\footnote{We rescale the Abbott-Deser mass by the factor $4\pi/\omega_\iota$ so that the coefficient of the $1/r$ order term in the inverse of the $r$-$r$ component of the metric becomes $2 m_{\rm AD}$ in a coordinate based on the areal radius ${\cal R}_A$ like Eq.~\eqref{fiota}, as in the asymptotically global AdS cases where $m_{\rm AD}$ coincides with $\mu_{\rm AD}$.}.

The Geroch monotonicity can be derived in the same way as that in Sec.~\ref{gerochmon} ,
 but for cases of asymptotically AdS we need to assume the global flow of the inverse mean curvature flow%
~\footnote{
The weak inverse mean curvature flow and the Geroch monotonicity under the weak flow are discussed in Ref.~\cite{Neves}, and it has been clarified that in some case the weak flow does not work. If the mass is nonpositive, the weak flow exists~\cite{Lee:2015xha}.
}.
The contributions of the cosmological constant are canceled out, and finally, 
the same inequality as Eq.~\eqref{mon.geroch3} is obtained but the overall constant $1/(32\pi^{3/2})$ is replaced by $2/(4\omega_\iota)^{3/2}$.
The energy outside $\sigma_0$, $m_{\rm ext}$, also has the same form as Eq.~\eqref{mextdef}, but the areal radius is defined based on $\omega_\iota$ as 
\begin{eqnarray}
{\cal R}_A := \sqrt{\frac{A}{\omega_\iota}},
\end{eqnarray}
and overall factor $2\pi$ in Eq.~\eqref{mextdef} is replaced with $\omega_\iota/2$. 
For other variables, by replacing $\pi$ in the discussion of asymptotically flat cases by $\omega_\iota/4$, 
most of the discussions in the previous sections are still valid. 

For a refined AGPS $\sigma_0$, a modified version of Eq.~\eqref{refinedAGPSonsigma}, i.e., 
\begin{align}
	\left(1+\frac{4}{3}\alpha\right)\int_{\sigma_0}dAk^2\le\frac{4}{3}\int_{\sigma_0}dA\left(-k{\cal D}^2\left(k^{-1}\right)+\frac{1}{2}{}^{(2)}R\right)\le\frac{8\pi}{3} \chi 
\end{align}
 is satisfied, where $\chi$ is the Euler characteristic of $\sigma_0$. For other AGPSs, the estimation of the integral of $k^2$ requires the Einstein equation and thus involves the cosmological constant. This contribution gives the difference from the final inequalities for the other AGPSs. 
For the refined AGPS, the extended Geroch energy on $\sigma_0$ becomes
\begin{eqnarray}
	E(0)&=&\frac{A_0^{1/2}}{\left(4 \omega_\iota\right)^{3/2}}\int_{\sigma_0}\left(2{}^{(2)}R-k^2-\frac{4\Lambda}{3}\right)dA \nonumber \\
	&\ge&  \frac{1+2\alpha}{3+4\alpha} \iota {\cal R}_{A0} - \frac{\Lambda}{6}{\cal R}_{A0}^3,
\end{eqnarray}
where we have used the fact that $\iota \omega_\iota = 2 \pi \chi$. 
Then, we can derive the following theorem:
\begin{theorem}
	\label{thm..refiend-adsagps}
              {\it	Let $M$ be an asymptotically locally AdS spacetime whose metric asymptotically approaches to Eq.~\eqref{metaAdS}, and $\Sigma$ in $M$ be an asymptotically locally AdS spacelike maximal hypersurface foliated by the inverse mean curvature flow $\lbrace \sigma_y \rbrace_{y \in {\mathbb R}}$ with $\sigma_y \approx \omega_\iota$, where $\sigma_0$ is a refined AGPS.
	Assuming that 
	no electromagnetic charges exist outside $\sigma_0$, we have an inequality for $\sigma_0$,
\begin{equation}
	\label{cragpsAdS}
	m_{\rm AD}\ \ge m_{\rm AD}-m_{\rm ext}\ge\frac{1+2\alpha}{3+4\alpha}\iota{\cal R}_{A0}- \frac{\Lambda}{6} {\cal R}_{A0}^3 +\frac{1}{2}\frac{q^2}{{\cal R}_{A0}}+\frac{\bar{J}_{\rm min}^2}{{\cal R}^3_{A0}},
\end{equation}
where $m_{\rm AD}$ is the  rescaled Abbott-Deser mass.
}
\end{theorem}
We should comment on definition of $\bar J_{\rm min}$, which is the minimum of $\bar J_y$. 
The definition of $\bar J_y$ is the same as that in the asymptotically flat cases, as given in Eq.~\eqref{Jydef}. 
The definition stems from the Komar angular momentum (see Eq.~\eqref{Komar}). 
In asymptotically AdS spaces (with the axial symmetry), the Komar angular momentum is usually normalized as 
\begin{equation}
\int_{\sigma_y} v^a \phi_a dA = \frac{8 \pi J}{\Xi^2} \qquad \mbox{with} \qquad \Xi:= 1+ \frac{a^2 \Lambda}{3}.
\end{equation}
Although it might be better to normalize our angular momentum $\bar J_{y}$, we continue to use the definition of Eq.~\eqref{Jydef} here.

The difference from the asymptotically flat case of Eq.~\eqref{cragps} is a positive contribution of the term $- \Lambda {\cal R}_{A0}^3/6$ 
in the right-hand side. 
For the LAGPS-k and TAGPS-k, comparing with the asymptotically flat cases of Eqs.~\eqref{clagps-k} and \eqref{ctagps-k}, other than the term $- \Lambda {\cal R}_{A0}^3/6$, we have an additional contribution $\Lambda/[2(3+4\alpha)]$ and $\Lambda/[2(3+4\beta)]$, respectively, which stems from the estimate of the integral of $k^2$. 
For the LAGPS-r and TAGPS-r, the additional contribution of the estimate for the integral of $k^2$ cancels with $- \Lambda {\cal R}_{A0}^3/6$, and 
the final inequalities are the same as Eqs.~\eqref{clagps-r} and \eqref{ctagps-r}. 
The inequality for a refined AGPS is the strongest one, in the sense that the additional contribution is the largest, and thus the lower bound for the rescaled Abbott-Deser mass becomes the largest. 

Let us consider the concept of extremality
by studying the case where the contribution of the angular momentum $\bar{J}_{\rm min}^2$ is ignored. Then, Eq.~\eqref{cragpsAdS} becomes
\begin{equation}
\frac{2(1+2\alpha)}{3+4\alpha}\iota- \frac{\Lambda}{3} {\cal R}_{A0}^2 - \frac{ 2m_{\rm AD}}{{\cal R}_{A0}} +\frac{q^2}{{\cal R}_{A0}^2} \le 0.
\label{AdSext}
\end{equation}
The equality in this inequality gives the extremality condition for the refined AGPS, that is, 
on an asymptotically locally AdS hypersurface $\Sigma$ in an asymptotically locally AdS spacetime $M$ with the  rescaled Abbott-Deser mass $m_{\rm AD}$ and the electromagnetic charge $q=\sqrt{q_e^2+q_m^2}$, 
no refined AGPS with parameter $\alpha$ exists
unless this inequality is satisfied for any ${\cal R}_{A0}$ for a fixed $\alpha$. 
The meaning of the extremality is the same as that discussed in the previous section. 
Note that, for $\alpha \to \infty$, Eq.~\eqref{AdSext} becomes the extremality condition for minimal surfaces or black hole horizons.


\section{Summary and discussion}
\label{summary}

In this paper, we have refined the area inequalities for five types of AGPSs.
The obtained inequalities are generalizations of the Riemannian Penrose inequality including the contributions 
of electric and magnetic charges, angular momentum, gravitational waves and matters. 
An interesting implication of the obtained results is the possibility of constructing thermodynamic
relations for surfaces in weak gravity region. 
In particular, there may exist the relation to Wald's entropy formula, which includes the contributions of energy  
outside the horizon~\cite{Sorce:2017dst,Wald:1993nt,Iyer:1995kg,Hollands:2024vbe}. 
The discussion of Wald's entropy formula would shed light on the original motivation of the Penrose inequality, that is, its relation to the cosmic censorship conjecture.
These interesting problems are left as the future work. 

The obtained inequalities
have led to the concept of extremality of the AGPSs. The 
extremality conditions for AGPSs are given as the relations between the ADM mass and  the electromagnetic charges
as presented in Sec.~\ref{cor}. 
The extremality conditions for black holes have recently been used in the discussion of the swampland program~\cite{Vafa:2005ui,Arkani-Hamed:2006emk}. 
It is expected that the extremality conditions for AGPSs are also applicable to the swampland program to give new constraints on quantum gravity. 

Our analysis is applicable to asymptotically locally AdS spacetimes as well.
In Sec.~\ref{AdS}, the inequalities and the extremality conditions in the case of asymptotically locally AdS spacetimes have been derived.
It would be interesting to explore their meanings in the context of the AdS/CFT correspondence.

\acknowledgments
K. I. would like to thank D. Yoshida for fruitful discussions.
K. I. and T. S. are supported by Grant-Aid for Scientific Research from Ministry of Education, Science, Sports and Culture of Japan (JP21H05182). 
K. I., H. Y. and T. S. are also supported by JSPS(No. JP21H05189). 
K. I. is also supported by JSPS Grants-in-Aid for Scientific Research (C) (JP24K07046). 
T. S. is also supported by JSPS Grants-in-Aid for Scientific Research (C) (JP21K03551), Fund for the Promotion of Joint International Research (JP23KK0048)
and Grant-in-Aid for Scientific Research (A) (JP24H00183). 
H. Y. is in part supported by JSPS KAKENHI Grant Numbers JP22H01220, and is partly supported by MEXT Promotion of Distinctive Joint Research Center Program JPMXP0723833165.

  
\end{document}